\documentclass[aps,prl,twocolumn,showpacs,amsmath,amssymb,nofootinbib,eqsecnum, preprintnumbers]{revtex4-1} 

\usepackage{color}
\newcommand{\be}{\begin{equation}}
\newcommand{\ee}{\end{equation}}
\newcommand{\ba}{\begin{eqnarray}}
\newcommand{\ea}{\end{eqnarray}}
\newcommand{\nn}{\nonumber}

\def\ben{\begin{equation}}
\def\een{\end{equation}}
\def\half{\frac{1}{2}}
\def\bea{\begin{eqnarray}}
\def\eea{\end{eqnarray}}

\def\nn{\nonumber}\def\p{\partial}
\begin{document}
\newcount\hour \newcount\minute
\hour=\time  \divide \hour by 60
\minute=\time
\loop \ifnum \minute > 59 \advance \minute by -60 \repeat
\def\nowtwelve{\ifnum \hour<13 \number\hour:% 		% supresses leading 0's
                      \ifnum \minute<10 0\fi%		% so add it it
                      \number\minute
                      \ifnum \hour<12 \ A.M.\else \ P.M.\fi
	 \else \advance \hour by -12 \number\hour:% 	% supresses leading 0's
                      \ifnum \minute<10 0\fi%		% add it in
                      \number\minute \ P.M.\fi}
\def\nowtwentyfour{\ifnum \hour<10 0\fi% 		% need a leading 0 
		\number\hour:% 				% supresses leading 0's
         	\ifnum \minute<10 0\fi% 		% add it in
         	\number\minute}
\def \now {\nowtwelve}

\begin{flushright}
\hfill UPR-1257-T
\end{flushright}

\title{Exact quasi-normal modes for the near horizon Kerr metric}

\author{M. Cveti\v c}
\email{cvetic@physics.upenn.edu}
 \affiliation{Department of Physics and Astronomy,
 University of Pennsylvania, Philadelphia, PA 19104, USA 
\\ \& Center for Applied Mathematics and Theoretical Physics,
University of Maribor, Maribor, Slovenia}
\author{G.W. Gibbons}
\email{g.w.gibbons@damtp.cam.ac.uk}
\affiliation{DAMTP, University of Cambridge, Wilberforce Road, Cambridge CB3 0WA, UK\\ 
\& Department of Physics and Astronomy,
 University of Pennsylvania, Philadelphia, PA 19104, USA }

\date{\today}

\begin{abstract}
We study the quasi-normal modes of a massless scalar field
in a  general sub-extreme  Kerr background
by exploiting  the  hidden  $SL(2,\mathbb{R} )\times SL(2, \mathbb{ R}) \times SO(3)$ symmetry of the  subtracted geometry approximation. 
This  faithfully models
the near horizon geometry but locates the black hole in a confining
asymptotically conical  box analogous to the anti-de-Sitter
backgrounds used in string theory. There are  just  two series of modes, given in terms of hypergeometric functions
and spherical harmonics, reminiscent of the left-moving and right-moving degrees in string theory:
one is overdamped, the other  is 
underdamped and  exhibits rotational splitting. 
The remarkably simple  exact formulae for the complex frequencies
would in principle allow the determination of the mass and angular
momentum  from observations of a black hole.
No black hole bomb is possible
because the Killing field which co-rotates with the horizon
is everywhere timelike outside the black hole.

\end{abstract}


\preprint{}

\maketitle

\section{Introduction}
Attempts to understand the quantum mechanics of black holes
in String/M-theory have led to the construction of a large number
of charged, rotating black hole solutions of supergravity theories.   
These theories admit several generalised Maxwell fields and 
in addition more than one scalar field.  
The standard neutral Kerr solutions  used to model astrophysical black holes
 belongs to a   family of such  metrics \cite{Cvetic:1996kv} 
which take  the form:\footnote{We set Newton's constant and the speed of light equal 1.}
\ben
d  s^2  = -  \Delta^{-1/2}_0  G  
( d{ t}+{ {\cal  A}})^2 + { \Delta}^{1/2}_0 
(\frac{d r^2} { X} + 
d\theta^2 + \frac{ X}{  G} \sin^2\theta d\phi^2 ),\label{metricg4d}
\een
with 
\bea
 X & =&  r^2 - 2 m r +  a^2 \, ,\nn\\
G & = & r^2 - 2 mr + a^2 \cos^2 \theta  \,.\nn\\
\eea
Remarkably it has been found  that for the entire family,  
the massless scalar wave equation is separable
and the solutions  expressible  in terms of spheroidal functions
of $\theta$  \cite{Cvetic:1997uw,Cvetic:1997xv}. 
The general form for $\Delta_0(r)$ and $\cal A$ is complicated but  
in the astrophysically relevant Kerr case   
it simplifies:  
\bea
 \Delta _0 &=&  r^4   +2 r^2a^2 \cos ^2 \theta  +  a^4 \cos^4\theta\, , 
\nn \\
{\cal A} &=& \frac{2mar \sin ^2 \theta }{G} d \phi \, , 
\eea
and the radial function may be expressed in terms of solutions of  
a confluent form of Heun's
equation which has two regular singular points   and an 
irregular singular point at infinity.

In \cite{Cvetic:2011dn} 
it was discovered, in the  much more general context of these
multi-charged black hole solutions \cite{Cvetic:1996kv} of the 
so-called STU model,
that making the replacements\footnote{In (\ref{deltas}), $\Pi_c=\Pi_{i=1}^4
\cosh\delta_i$ and $\Pi_s=\Pi_{i=1}^4\sinh\delta_i$, where the  four boost parameters $\delta_i$ parameterize the four charges $Q_i=2m\sinh\delta_i\cosh\delta_i$ of the original four-charge rotating solution. }

\bea
\Delta_0 \rightarrow  \Delta&= & (2m)^3r(\Pi_c^2-\Pi^2_s) +(2m)^2\Pi_s^2\nn\\
&-&(2m)^2 (\Pi_c-\Pi_s)^2 a^2 \cos^2 \theta \,    \label{deltas}
\eea
in (\ref{metricg4d})  hence reducing the highest  
power of $r$ in $ \Delta_0$
renders  the irregular singular point at infinity regular, allowing
solutions in terms of hypergeometric functions.    Moreover the massless scalar wave equation is now separable
in terms of ordinary spherical harmonics, rather than the complicated
spheroidal functions needed for the Kerr solution. In addition, while
the  so-called  ``subtracted metric''  remains  non-spherically symmetric, 
the massless scalar wave equation    
exhibits a hidden $SL(2,\mathbb{R} )\times SL(2, \mathbb{ R}) \times SO(3)$
symmetry.  Furthermore the areas of the outer and inner
horizons, their angular velocities $\Omega_\pm$ and  and their surface
gravities  $\kappa_\pm$ are unchanged by this replacement, 
preserving the  local geometry and thermodynamic properties of the metric. 
Thus in the special case of the Kerr solution
the associated  subtracted  metric faithfully models the near horizon
environment of a general neutral non-extreme  Kerr black hole.\footnote{For a different approach to the massless wave equation  in the Kerr background, valid at wave lengths large compared to the horizon scale, see \cite{Castro:2010fd}.}  
The aim of this letter is to exploit this fact 
in order to study analytically 
its  quasi-normal modes in this near horizon approximation.

This enormous simplification was
however bought at the cost that the
subtracted metric  is no longer asymptotically flat but rather
asymptotically conical:
\ben
ds ^2 \approx - \bigl( \frac{R}{R_0}\bigr ) ^6 dt^2+16 dR^2 + R^2 \bigl(d \theta ^2 + \sin ^2 \theta d \phi ^2  \bigr )\, ,  \label{conical}     
\een
where $R= (8m^3r ) ^{\frac{1}{4} }$, $R_0=(2m^3)^{\frac{1}{3}}$ and  
$-g_{tt}$ increases  as radial distance $4R $ to the power six
corresponding to a logarithmically increasing gravitational
potential. 
This suggests that
the subtracted metric  describes a black hole confined in 
a box in a way which is analogous to the behaviour of a Kerr-anti-De-Sitter
black hole \cite{Cvetic:2012tr}.  One expects the subtracted geometry 
to be an excellent  approximation to the exact
Kerr solution  near the horizon but clearly deviates
from it considerably at large distances. 
Except in the purely radially directed case,
the asymptotic  conical (\ref{conical})   geometry admits 
no  null geodesics which reach infinity \cite{Cvetic:2012tr} and so   
it is best viewed as a good  approximation well inside $r=3m$,
the radius of the photon sphere in the Schwarschild case.    
  
The massless scalar wave equation in the
Kerr-Anti-De-Sitter background has been the subject of enormous
interest of late because  of its relevance to the AdS/CFT 
correspondence. Of  particular concern  has been the behaviour of
quasi-normal modes and stability issues. In the Kerr-anti-De-Sitter
case it has been found \cite{Cardoso:2004nk} that for a suitable range of parameters
the  system is unstable,  behaving like a   ``black hole bomb'' \cite{Press:1972zz}.
Of necessity these investigations have been numerical since
neither the radial functions nor the spheroidal functions
admit simple analytic expressions.

By contrast for the subtracted Kerr  metric, the quasi-normal modes
may be found exactly using standard properties of spherical
harmonics and hypergeometric functions.  We find, using the results of 
\cite{Cvetic:1997uw,Cvetic:1997xv,Cvetic:2011dn}, that
there are two  discrete series of  solutions, each member  being
labelled by by three  integers $l=0,1,\dots , n = \pm l , \pm (l-1), \dots
$ and $N_R,N_L =0,1, \dots$,   
and having  time dependence
\be
e^{ -\frac{1}{4m} ( l+1+N_L ) t} \, \quad \hbox{or}\quad
e^{- \frac{\sqrt{m^2-a^2}}{4 m^2 }( l+1+N_L ) t} 
e^{-i n\frac{a}{2m^2} t }\, .  
\label{modes}
\ee
Both sets of modes decay exponentially in amplitude: one is overdamped, the other  is 
underdamped and  exhibits rotational splitting. 
The absence of an instability due to super-radiant behaviour
may be understood  as a consequence of the fact that the 
Killing vector $l^+=\p_t + \Omega_+ \p_\phi$, 
which 
coincides on the horizon with its null generator, 
is timelike everywhere outside the horizon.
By a result of Hawking and Reall \cite{Hawking:1999dp},  this 
is sufficient to preclude super-radiance instabilities. 

As expected,  the damped oscillatory series of modes exhibit
rotational splitting, i.e. dependence on the angular quantum number
$n$,  due to gravito-magnetic or frame dragging effects analogous
to Zeeman splitting in atomic physics or the free oscillations
of the earth \cite{Ness,Benioff}. The extraordinarily simple form
of (\ref{modes}) allows in principle their use 
in determining  the angular momentum $a$ and the $m$ from observational data. 

In what follows we 
shall provide some further details of our results.

\section{The metric}
For Kerr solution  one has to set  in (\ref{deltas})   $\Pi_c=\Pi_{i=1}^4
\cosh\delta_i=1$ and $\Pi_s=\Pi_{i=1}^4\sinh\delta_i=0$, i.e.   all  $\delta_i=0$. 
The subtracted Kerr metric   factor (\ref{deltas})  is
\be
\Delta = (2m)^3 r -(2m)^2  a^2 \cos^2 \theta   \,.  
\ee
At the inner and outer horizon:
 \be 
r_\pm =  m\pm \sqrt{m^2 - a^2}~,
\ee
the angular velociites and surface gravities:  
\ben
\Omega _\pm =  \frac{a}{ 2m ( m \pm  \sqrt{ m^2-a^2 }) }, \,   \kappa_\pm = \frac{\sqrt{m^2 - a^2}}  {2m ( m \pm \sqrt{ m^2-a^2 })  }\, ,
\een
remain the same as for the Kerr black hole. Note that\footnote{This is a general properly of  the multi-charged STU black hole solutions
\cite{Cvetic:1996kv}, observed already in \cite{Cvetic:1997uw,Cvetic:1997xv}.}

\be \frac{\Omega_+}{\kappa_+}=\frac{\Omega_-}{\kappa_-}=\frac{a}{\sqrt{m^2-a^2}}\, . \label{pme}
\ee

The subtracted Kerr metric itself  can be cast  in the following remarkable form:

\bea
ds^2 &=& 
 \sqrt{\Delta} \frac{X}{F^2}
  \bigl(- dt ^2 + \frac{F^2 dr^2} {X^2 } \bigr )\nn\\
  &+&
  \sqrt{\Delta} d\theta ^2 + \frac{F^2 \sin ^2 \theta}{ \sqrt{\Delta}  }  (d \phi + W d t) ^2  \, , \label{metricn}
\eea
where 
\be X=r^2-2mr +a^2,  W= -\frac{2mar}{F^2},  F^2=(2m)^3 r-(2m)^2a^2, 
\ee
depend only on the radial coordinate $r$.
Note, $W(r_\pm)=-\Omega_\pm$ and $F(r_\pm)=\frac{r_+-r_-}{2\kappa_{\pm}}$.

\section{Kruskal-Szekeres Coordinates}
We now construct Kruskal-Szekeres type  coordinates to cover
the outer horizon which allow us to identify suitable
boundary conditions there.\footnote{One can analogously construct  Kruskal-Szekeres type  coordinates to cover
the  inner horizon  region.} At infinity the 
appropriate boundary condition is boundedness of the solution. 
The construction of Kruskal-Szekeres  coordinates is in fact considerably simpler than that
used for the Kerr solution \cite{Carter1,Carter2}.

Eddington-Finkelstein null coordinates $u,v$ take  the form
\ben
u=t-r^\star  \,,\qquad v=t+ r^\star  
\een 
and satisfy
\ben
g^{\alpha \beta } \p _\alpha u \p_  \beta u  = 0 =
 g^{\alpha \beta } \p _\alpha v \p_  \beta v \, .
\een 
Using the fact that the Hamilton-Jacobi equation is separable, we find that 
\ben
 r^\star = \int ^r  \frac{F  dr}{X }\,,\een
and we define an ingoing angular coordinate 
\ben
 \phi_+ =  \phi - \Omega_+ t  \,, 
\een
which is constant along the orbits of the co-rotating Killing vector $l^+$
\ben
l^+  \phi_+ = (\p_t + \Omega _+ \p_\phi) \phi_+ =0\,. 
\een
Note that the above results are manifest for the metric written as (\ref{metricn}).

Kruskal-Szekeres coordinates 
\be U=-e^{-\kappa_+u}\, , \quad V=e^{\kappa_+v}\, , 
\ee
imply
\be
\frac{ 2 \kappa _+ F dr}
{X } = dUU^{-1}+ dVV^{-1}\, , \
2 \kappa_+ dt
= dVV^{-1}-dUU^{-1}\, .
\ee
In the vicinity of outer horizon, $r\sim r_+$, one obtains $-UV\sim (r-r_+)$, where we have employed $F\sim\frac{r_+-r_-}{2\kappa_+}$. One can then check explicitly that in  coordinates $(U,V, \phi_+, \theta)$,
 the metric (\ref{metricn}) is non-singular and  analytic.   
 Moreover, the length-square of the co-rotating Killing vector: 
\be
g^{\alpha \beta }l^+ _\alpha   l^+ _\beta =-\Delta^{-\frac{1}{2}}\bigl[X+
\frac{a^2\sin^2\theta(r_+-r_-)(r-r_+)}{r_+^2}\bigr]
\, , \ee
is  manifestly negative for  all $ r>r_+$.

\section{Solutions of the Wave equation}
The separated solutions of the massless scalar wave equation 
are  of the form $e^{-i\omega t} e^{in \phi} P^n_l(\theta) \chi(x)$ 
where  $P^n_l(\theta)$ is an associated Legendre 
polynomial and  \cite{Cvetic:1997uw,Cvetic:1997xv}  
\ben
x = \frac{r - \frac{1}{2}(r_+ + r_-)} { r_+ - r_-}~,
\een
is designed so that the  two horizons $r_\pm$ are at   
$x=\pm \frac{1}{ 2}$ . The overall scale of the black hole is set by $r_+ + r_- = 2m$.
The radial wave equation  takes the form \cite{Cvetic:2011dn}
\bea
&&\Bigl[ \frac{\partial}{\partial x} (x^2 - 
\frac{1}{4})\frac{\partial}{\partial x} +  
\frac{1}{4(x-\half )}\bigl( \frac{\omega}{\kappa_+} - 
n\frac{\Omega_+}{\kappa_+} \bigr)^2 \nn \\
  - &&\frac{1}{4( x+\half)}
 \bigl(\frac{\omega}{\kappa_-}  -
 n \frac{\Omega_+}{\kappa_+} \bigr)^2 -   l(l+1) \Bigr] \chi(x)=0.
\eea
Note that in the equation we already employed the symmetry of the ratios (\ref{pme}).

Solutions which are ingoing on the future horizon
must be regular at $U=0$ in Kruskal-Szekeres coordinates and
this implies that \cite{Cvetic:1997uw,Cvetic:1997xv}   
\bea
 &&\chi(x)= (x+\half)^{- (l+1) }
\bigl( \frac{x-\half}{x+\half } \bigr ) ^{ -i \frac{\beta_H }{4 \pi }(\omega - n \Omega _+) }\nn \\
&&F ( l+1-  i  \frac{\beta_R \omega -2n \beta _H \Omega _+}{2 \pi},\nn \\
&& l+1-  i  \frac{\beta_L\omega }{2 \pi}, 1-   i \frac{\beta_H (\omega - 
n \Omega _+ ) }{2 \pi }
; \frac{x-\half}{x+\half } ) \,,  \eea
where 
\ben
\frac{\beta _H}{2 \pi} = \frac{1}{\kappa_+} \,, \quad 
 \frac{\beta _R}{2 \pi } = \frac{1}{\kappa_+} + \frac{1}{\kappa _-} \,,\quad 
\frac{\beta _L}{2 \pi } = \frac{1}{\kappa_+} - \frac{1}{\kappa _-}\,.  
\een
Near the outer horizon  $r^\star \rightarrow - \infty,\,
\frac{x-\half}{x+\half } \rightarrow e^{2 \kappa_+ r^\star } 
$  and so 
\bea
&& \chi(x) \approx e^{-i  (\omega - 
n \Omega _+)  r^\star} \nn\\
&&F( l+1-  i  \frac{\beta_R \omega -2n \beta _H \Omega _+  }{2 \pi},\nn\\
&& l+1-  i  \frac{\beta_L\omega }{2 \pi}, 1-   i \frac{ \beta_H(\omega  - n \Omega _+)  }{2 \pi }
; e^{2 \kappa _+ r^\star}  )\, .
\label{ingoing} \eea

In Kruskal-Szekeres coordinates therefore
\ben
e^{-i\omega t} e^{in \phi} \chi (x) \approx e^{in \phi_+ } 
 V^{i\frac{\omega - n \Omega_+}{\kappa _+ } }  ( 1 + \dots ) \, ,   
\een
where the ellipsis denotes a power series in $UV$
which is convergent in a neighbourhood of the future horizon 
$U=0$. 

At large $x$  \cite{Cvetic:1997uw,Cvetic:1997xv}
\bea
&&\chi(x) \approx  x ^{-(l+1)}
\frac{
\Gamma (1 -i  \frac{\beta_H (\omega -n  \Omega _+) }{2 \pi }) \Gamma (-2l-1 ) }
 {\Gamma (-l -i \frac{\beta_L \omega}{2 \pi})
 \Gamma (-l -i  \frac{\omega \beta_R - 2n \beta _H \Omega _+ }{2 \pi}  )
}\nn \\
&+& x^l  
\frac{\Gamma (1 -i  \frac{\omega \beta_H (\omega -n\Omega _+)  
}{2 \pi }) \Gamma (2l  +1 )} 
{\Gamma ( l+1 - i  \frac{\beta_L\omega}{2 \pi}) 
\Gamma (l+1 -i   \frac{\omega \beta_R - 2n \beta _H \Omega _+ } {2 \pi}) 
} \, . 
\eea
In order that $\chi(x)$ be finite at large $x$, we must set  
\bea
&&i \omega  \frac{\beta_L}{2 \pi} = l+1 +N_L\, , \nn\\
\hbox{or}\quad
&&i   \frac{\omega \beta_R - 2n \beta _H \Omega _+}{2 \pi}
=l+1 +N_R \,, \eea
where $N_{L,R}=0,1,\dots$ This gives remarkably simple  formulae for the frequencies of the quasi-normal modes:
\bea
&&\omega=-\frac{i}{4m}(1+l+N_L)\, , \nn\\
\hbox{or}\quad &&\omega=-\frac{i\sqrt{m^2-a^2}}{4m^2}(1+l+N_R)+ \frac{a}{2m^2}n\,.
\eea
  Both  frequencies result in damped modes, with the underdamped branch exhibiting oscillatory behavior and the  damping absent  in the extremal limit $a\to m$. The specific asymmetry  in the frequencies of the two branches is due to  (\ref{pme}).The two branches of modes are reminiscent of the left-moving and right-moving modes in string theory, thus potentially relating this subtracted geometry  microscopically to a  dual two-dimensional conformal field theory description \cite{Cvetic:2011dn} and hence the reference to $L$ and $R$ in the preceding formulae.  

\section{Conclusion and future prospects}
In this letter we have obtained exact formulae for the
quasi-normal modes of the Kerr solution in the subtracted geometry
approximation. It would be interesting to compare our results
with numerical calculations of the massless wave equation
for initial date with support confined to a small neighbourhood of the horizon.
   
The subtracted geometry is a solution \cite{Cvetic:2012tr}
 of  the equations of motion for the Lagrangian of N=2 supergravity coupled to three vector super-multiplets, often referred to as  the  STU model.  Its lift to five-dimensions corresponds to $AdS_3 \times S^2$ \cite{Cvetic:2011dn,Cvetic:2012tr}.   By taking a scaling limit of   Melvin STU black holes in this model, a subtracted geometry with the magnetic field  parameter $\beta$ for the Kaluza-Klein  gauge  field was obtained  \cite{Cvetic:2013roa}. The lift of this geometry again results in $AdS_3 \times S^2$  where the azimuthal angle is shifted by a coordinate transformation in the circle direction $z$ as $\phi \to \phi - \beta  z$.  As a consequence, the wave equation for the massless minimally coupled scalar in this background separates and allows for an explicit  analysis of the quasi-normal modes. Details will be presented elsewhere \cite{Cvetic:2013lll}.

 {\bf Acknowledgements} We would like to thank Zain Saleem for discussions and collaboration on related topics.
 This research is supported in part by the DOE grant DE-SC0007901,  the Fay R. and Eugene L. Langberg Endowed
Chair (M.C.) and the Slovenian Research Agency (ARRS) (M.C.).

%\newpage
%\bibliographystyle{apsrmp}
%\bibliography{Bibliography_Toda}

\providecommand{\href}[2]{#2}

\begingroup\raggedright

\end{document}